\documentclass[aps,prl,twocolumn,10pt,superscriptaddress,amsmath]{revtex4-1}

\usepackage[utf8x]{inputenc}

\usepackage{amsmath}
\usepackage{amsfonts}
\usepackage{amssymb}
\usepackage{graphicx}

\usepackage[bookmarks=true,colorlinks=true,citecolor=blue,linkcolor=red,urlcolor=magenta]{hyperref}

\begin{document}

\title{Comment on "Observation of electron-antineutrino disappearance at Daya Bay"}

\author{V.D. Rusov}
\email{siiis@te.net.ua}

\author{V.A. Tarasov}
\author{S.A. Chernegenko}
\author{V.P. Smolyar}
\affiliation{Department of Theoretical and Experimental Nuclear Physics, Odessa National Polytechnic University, Odessa, Ukraine}

\maketitle

In Letter by An \textit{et~al.} \cite{bib-1} Daya Bay measurements of non-zero value for the neutrino mixing angle $\theta_{13}$ with a significance of 5.2 standard deviations were reported. The value of $\sin ^2 \theta_{13}$ was determined with a $\chi ^2$ constructed with pull terms accounting for the correlation of the systematic errors. The survival probability used in the $\chi ^2$ was \cite{bib-1}

\begin{align}
%\begin{equation}
P_{sur} &= 1 - \sin ^2 2\theta_{13} \sin ^2 \left( \frac{1.267 \Delta m^2 _{31} L}{E} \right) - \nonumber \\
&- \cos ^4 \theta_{13} \sin ^2 2\theta_{12} \sin ^2 \left( \frac{1.267 \Delta m^2 _{21} L}{E} \right),
\label{eq1}
%\end{equation}
\end{align}

\noindent where the parameters of mixing $\Delta m ^2 _{21} = 7.59 \cdot 10^{-5} ~eV^2$, $\sin ^2 2 \theta _{12} = 0.861$ were obtained within the KamLAND–experiment \cite{bib-2}, taking into account the nonzero nuclear georeactor hypothesis. However, it follows from the nuclear reactors physics that applying these parameters to (\ref{eq1}) is not quite correct, because the nonzero nuclear georeactor hypothesis used in KamLAND \cite{bib-2} and Borexino \cite{bib-3} experiments, is not physically complete. The hypothesis incompleteness is associated with a total  neglect of a substantial nuclear georeactor nonstationarity under the extreme conditions (temperature $\sim$5000-6000 K and pressure $\sim$340-360 GPa at the Earth core), which implies an absolute ignoring of a high nonequilibrium of georeactor antineutrino spectra. Failing to take into account the georeactor antineutrino spectra nonequilibrium (caused by the disturbance (10-15\% and more \cite{bib-4}) of the secular equilibrium of all without exception fission products) may eventually lead to the incorrect neutrino parameters of mixing determination in KamLAND-experiments \cite{bib-5}. Let us show why it is so.
It is known that a Herndon fast reactor \cite{bib-6,*bib-6a} was used in the framework of a nuclear georeactor hypothesis in KamLAND and Borexino experiments. The heat power evolution of such reactor is described by the following nuclear transmutations chain within the uranium-plutonium cycle:

\begin{equation}
^{238} U \left( n, \gamma \right) \rightarrow ^{239}U \xrightarrow[]{\beta^{-}} {^{239} Np} \xrightarrow[]{\beta^{-}} \left\lbrace 
\begin{matrix}
^{239} Pu \xrightarrow[]{\alpha} ^{235}U \\
^{239} Pu (n,fission)
\end{matrix}
\right.
\label{eq2}
\end{equation}

\begin{figure}
\includegraphics[width=\linewidth]{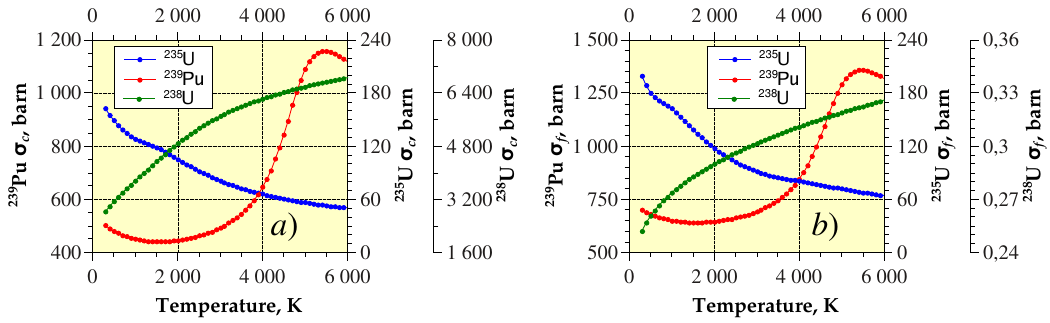}
\caption{Dependence of (a) capture cross-sections and (b) fission cross-sections for  $^{235}$U (blue), $^{238}$U (green), and $^{239}$Pu (red) averaged over the neutron spectrum on fuel medium temperature for limiting energy (3kT) of the Fermi and Maxwell spectra. A neutron spectra averaging procedure was applied for the concentrational fuel composition of the nuclear georeactor discussed in~\cite{bib-8}.}
\label{fig1}
\end{figure}

Omitting a number of assumptions and suppositions given in these papers, it is enough to consider only one key parameter for our purpose – the so-called infinite multiplication factor $k_{\infty}$, which characterizes exhaustively the physics of the nuclear burning process in such reactors. It is easy to show \cite{bib-7} that the following condition must be satisfied:

\begin{equation}
k_{\infty} = \frac{\eta^{(5)} \sigma^{(5)}_a \overline{N} ^{(5)} + \eta^{(8)} \sigma^{(8)}_a \overline{N} ^{(8)} + \eta^{(9)} \sigma^{(9)}_a \overline{N} ^{(9)}}{\sigma^{(5)}_a \overline{N} ^{(5)} + \sigma^{(8)}_a \overline{N} ^{(8)} + \sigma^{(9)}_a \overline{N} ^{(9)}} = 1
\label{eq3}
\end{equation}

\noindent in order to maintain the criticality in 3-component system (\ref{eq2}). Here $\sigma _a ^{(i)}$ is a microscopic absorption cross-section of $^{235}$U (i$\equiv$5), $^{238}$U (i$\equiv$8) and $^{239}$U (i$\equiv$9), $\eta ^{(i)}$ – neutrons produced per absorbed neutron, $\overline{N} ^{(i)}$ is the equilibrium atomic particle density.

Obviously, as the opposite to technical reactor physics we have to note here (See~(\ref{eq3})) that the infinite multiplication factor is controlled by the absolute value of the critical neutron flux \cite{bib-7}. A very important conclusion follows immediately from said above: the inevitable variations of the neutron flux driven by the natural thermal variations of the components concentrations and the corresponding microscopic fission and capture cross-sections, will definitely produce a high nonstationarity of the Herndon fast reactor.

To illustrate the said above, we show on Fig.~\ref{fig1} the dependencies of capture and fission cross-sections on the fissile medium temperature for the main nuclear fuel components \cite{bib-5}, that were not taken into account in the papers by Herndon \cite{bib-6,*bib-6a}. It is a consequence of the fact that the SAS2 analysis sequence, which is a base for the numerical simulation calculations of the composition and heat power evolution for the nuclear georeactor of Herndon type \cite{bib-6,*bib-6a}, simply does not contain (just like any other modern reactor simulation code) the data about the cross-section profile in the mentioned temperature range.

%\begin{figure}
%\includegraphics[width=\linewidth]{fig1.eps}
%\caption{Dependence of (a) capture cross-sections and (b) fission cross-sections for  $^{235}$U, $^{238}$U, and $^{239}$Pu averaged over the neutron spectrum on fuel medium temperature for limiting energy (3kT) of the Fermi and Maxwell spectra. A neutron spectra averaging procedure was applied for the concentrational fuel composition of the nuclear georeactor discussed in~\cite{bib-8}.}
%\label{fig1}
%\end{figure}

A natural conclusion would be that ignoring nonequilibrium properties of georeactor neutrino spectra originating from the nonstationary nuclear georeactor operation mode or, in other words, description of the experimental effective neutrino spectrum by the regular equilibrium neutrino spectra either of individual nuclides or their mixture can result in serious mistakes in fitting the total experimental neutrino spectrum.

In summary, we believe that since the nonzero nuclear georeactor hypothesis used in KamLAND and Borexino experiments is not physically complete, it automatically gives rise to an unavoidable problem of taking into account the nonequilibrium geoneutrino spectrum of the nuclear burning process within the framework of nonzero nuclear georeactor $\chi^2$-hypothesis \cite{bib-4}. Thus a direct applying of the KamLAND parameters of mixing to Daya Bay-experiment (in contrast to Double Chooz Experiment \cite{bib-9}) will produce a high uncertainty level and, as a result, a substantial distortion of the actual value of the neutrino mixing angle $\theta_{13}$.

\begin{acknowledgements}
This work is partially supported by EU FP7 Marie Curie Actions, SP3-People, IRSES project BlackSeaHazNet (PIRSES-GA-2009-246874).
\end{acknowledgements}

\bibliographystyle{apsrev4-1.bst} 
\bibliography{Comment-DayaBay}

\end{document}